\documentclass[conference]{IEEEtran}

\usepackage{url,comment,cite}
\usepackage{amsmath,amssymb,amsfonts,bbm,dsfont}
\usepackage{graphicx}
\usepackage{epsfig,epsf,psfrag,textcomp,tabularx}
\usepackage[usenames,dvipsnames]{xcolor}
\usepackage{subfigure}
\usepackage{float}
\usepackage{stfloats,mathtools}
\usepackage[algoruled,medskip,dontprintsemicolon,linesnumbered,Algorithm]{algorithm}
\usepackage[noend]{algpseudocode}
\usepackage{autonum}
\usepackage{multirow}
\usepackage{bm}

\newcommand{\Bc}{\mathcal{B}}
\newcommand{\Gc}{\mathcal{G}}
\newcommand{\Vc}{\mathcal{V}}

\newcommand{\Cc}{\mathcal{C}}

\newcommand{\Kc}{\mathcal{K}}

\newcommand{\Zc}{\mathcal{Z}}

\newcommand{\Fig}[1]{Fig.~\ref{fig:#1}}

\newcommand{\Eq}[1]{(\ref{eq:#1})}

\begin{document}

\title{Graph-based Model for Beam Management\\
in Mmwave Vehicular Networks
}

\author{\IEEEauthorblockN{Zana Limani Fazliu$^\ast$, Carla Fabiana
    Chiasserini$^{\dagger,\S,\ddagger}$, Francesco Malandrino$^{\ddagger}$, Alessandro Nordio$^{\ddagger}$}
\IEEEauthorblockA{$\ast$ University of Prishtina, Kosovo
  $\quad$ $\dagger$ Politecnico di Torino, Italy $\quad$ $\S$ CNIT,  Italy $\quad$ $\ddagger$ CNR-IEIIT, Italy}
}

\maketitle

\begin{abstract}
Mmwave bands are being widely touted as a very promising option
for future 5G networks, especially in enabling such networks to
meet highly demanding rate requirements. Accordingly,
the usage of these bands is also receiving an increasing interest in the context of 5G 
vehicular networks, where it is expected that connected cars will
soon need to transmit and receive large amounts of data. 
Mmwave communications, however, require the link to be established
using narrow directed beams, to overcome harsh propagation
conditions. The advanced antenna systems enabling this also allow for
a complex beam design at the base station, where multiple beams of
different widths can be set up. In this work, we focus on beam
management in an urban vehicular network, using a graph-based approach
to model the system characteristics and the existing constraints. 
In particular, unlike previous work, we formulate the beam design problem as a maximum-weight
matching problem on a bipartite graph {\em with conflicts}, and then we solve it
using an efficient heuristic algorithm. Our results show that our
approach easily outperforms 
advanced methods based on clustering algorithms. 
\end{abstract}

\section{Introduction}
\label{sec:intro}

Due to the sheer amount of bandwidth available in millimeter-wave
(mmwave) bands, they are becoming an increasingly attractive choice
for various types of wireless networks. They are already included in
the standardization efforts for 5G \cite{zorzi-tutorial, 3gpp-tech},
and, as such, they are set to become an essential resource for
5G-enabled vehicular networks. Indeed, the increasing need for
on-board high-definition maps, their real-time updates, as well as the
data generated by the vehicles themselves, make conncted cars  
prime consumers and producers of network traffic, which can be catered to only by a substantial amount of bandwidth, readily available in mmwaves. 

However, due to the high operating frequency, these bands are known to experience harsh propagation conditions, and are highly susceptible to blockages, which has rendered them unusable until very recently \cite{rappaport-chanmodels}. Advances in large smart antenna systems, composed of many antenna elements, can overcome these problems by establishing communication links over narrow beams with high beamforming gains. In addition, these advanced systems can support several simultaneous beams of varying witdth \cite{kulkarni-hybridbf}. 

The fact that the communications are conducted over highly directed
beams 
 introduces both challenges and opportunities. Narrow beams, when efficiently used, can
 ensure a high degree of isolation from interference due to other
 ongoing communications, but they also
 significantly curb coverage and range of the mmwave base stations
 (gNBs). This means that beam management aspects in future networks
 will be no small feat. For this reason, it is not expected that
 mmwave service will be deployed in a stand-alone fashion, but rather
 in tandem with networks operating in sub-6 GHz frequencies, to
 alleviate shortcomings, especially during initial access and link establishment \cite{zorzi-lte}.

 A signifcant body of work has addressed this issue,
either in the context of initial access and link establishment, or
beam alignment/configuration and user association. 
In  \cite{perfecto-v2v} the authors present a framework that combines matching
theory and swarm intelligence to dynamically and efficiently
perform user association and beam alignment in a vehicle-to-vehicle communication network.
Methods aided by location information have been proposed in \cite{location-allerton}, as have methods which use information from road-side and vehicle sensors \cite {discover-5g, arxivali2019}. 
In our own previous work \cite{noi-wowmom19}, a method leveraging traffic regulating signals was proposed to alleviate the need for real-time beam realignment. 
An optimal mutli-user,
non-interactive beam alignment approach is proposed in
\cite{optimal-niba},  which however focuses  on a single-cell network.

%


In this work we
take a novel graph-based approach to the beam management problem.  In
particular, we address it by casting the joint beam design and user
association task as a {\em conflict-aware} matching problem in a
weighted bipartite graph, with the goal of ensuring broad coverage 
while maximizing the network data rate. 
 Few works in the literature  have
applied graph theory in general, to address beam management in mmwaves
\cite{mmwave-graph-icc, mmwave-graph-tvt}. 
In both \cite{mmwave-graph-icc, mmwave-graph-tvt},
the authors propse a graph-based approach to reduce the inter-cell
interference, whereby each mmwave cell/link is modelled as a vertex and the
edges between them respresent the mutual inferference caused. The goal
is to find a subgraph that minimizes the number of beam
collisions.



 To summarize, in this work we make the following main contributions:
\begin{itemize}

\item we formulate the beam design problem, i.e., the joint selection
  of the number, width and direction of beams,  as an
optimization problem, aiming at maximizing the rate of the covered
users, while respecting all practical constraints;

\item  we develop a graph-based  model of the mmwave system, which
  captures the most essential features. The optimization problem is
  thus turned into a problem of {\em bipartite weighted matching with
    conflicts}, which can be solved in linear time using heuristic algorithms. In particular, the introduction of {\em{conflicts}} within the graph model enables us to accomodate the practical constraints of mmwave communications;

\item  we evaluate our approach leveraging a large-scale
trace, including the real-world urban topology and realistic
vehicular mobility of Luxembourg City, Luxembourg, and compare it against a state-of-the-art cluster-based beam design approach. 
Our results show that the proposed scheme is able to provide better performance, in particular thanks to its ability to accomodate practical constraints into the solution mechanism.

\end{itemize}
Unlike previous work, which mainly addressed a single parameter at a
time, our approach jointly selects, for each gNB, three beam-design
parameters: the number of simultaneous beams, their width, and their
direction. 
Note that, while weighted bipartite matching  {\em without} conflicts
is a fairly well-studied  problem, the
same problem {\em with conflicts} has been much less
explored and applied.

The remainder of the paper is organized as follows. After
detailing the system
model and our vehicular network in Sec.~\ref{sec:system-trace},  
we formulate the optimization problem and introduce our graph-based
approach in Sec.~\ref{sec:problem}.  
In Sec.~\ref{sec:results},
through extensive simulations using real-world vehicular traces,
we present the performance evaluation. Finally, Sec.~\ref{sec:conclusion}
concludes the paper.

\section{The mmwave vehicular network}
\label{sec:system-trace}

\begin{figure}
\centering
\includegraphics[width=0.4\textwidth]{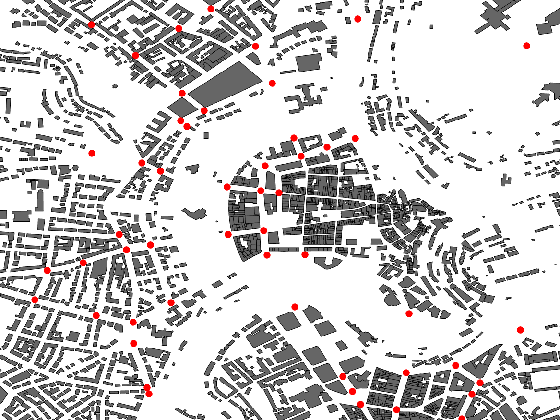}
\caption{\label{fig:scenario} Real-world scenario: Luxembourg city
  center. The red circles represent the locations of the traffic
  lights, i.e., of the gNBs.}
\end{figure}

We consider a realistic vehicular network in
an urban setting, based on real-world publicly available data for the
city of Luxembourg \cite{lust}. This data  contains sufficient
information about the topology of the city, the road layout (e.g.,
regulated intersections), as well as the mobility traces of around
14,000 vehicles traveling within the city center, accumulated over a
12-hour window.  Based on this data, we construct a scenario as
depicted in Fig.~\ref{fig:scenario}, in which a set of gNBs, denoted by $\Gc$,  are colocated with traffic lights to serve a set of
vehicles, i.e., the mmwave {\em{users}}, denoted by $\Vc$.

gNBs and  users are equipped with uniform planar array (UPA) antennas, composed of a grid of $N_t$ and $N_r$ antenna elements, respectively,
spaced by $\lambda/2$. Array antennas at the gNB can support up to $N$
beams simultaneously, limited by the number of available RF chains, while vehicles can use only one beam at a time.
For the mmwave communication between gNBs and users to be successful,
the beams need to be fully aligned at both the transmitting and
receiving end, or in the case of no line of sight (nLoS), directed in such a manner that the angle of arrival of the incoming waves coincides with the direction of the receiving beam. Moreover, the directivity and gain of a beam are inversely proportional to the width of the beam, i.e., the narrower the beam, the higher the gain. The number  and  width of the beams determine the range and coverage of the gNB, while the direction of the beam ensures alignment with the receiving beam, and isolation from interfering signals. It is clear therefore that the number, width, and directions of the beams are critical aspects that need to be addressed in a mmwave vehicular network. 

In this paper, we focus on downlink communications, although the work
can easily be extended to the uplink direction as well.  
Furthermore, multiple vehicles can be multiplexed within the same beam, using multiple access techniques.   
Coordinated multi point transmission (CoMP) is also supported, i.e., we assume that a vehicle can receive data through its single beam from several gNBs. 
Finally, 
to make the model more tractable, we divide the network area under
consideration into equal-sized square zones. We denote the set of
zones by $\Zc$ and make them sufficiently small so that their size is negligigle with respect to the footprint of any
beam. It follows that we can 
consider that all vehicles within a specific zone expericence the same
propagation and LoS conditions with respect to  the surrounding gNBs.

\section{Beam matching and user association}
\label{sec:problem}

We now focus on the main aspects of the mmwave vehicular network we
address and present our approach to overcome the existing hurdles. In
particular, Sec.\,\ref{sub:opt} formally formulates the optimization
problem for the user-gNB association,  stating its objective and
constraints. Then 
Sec.\,\ref{sub:graph} introduces our graph-based
model with constraints and a heuristic algorithm that effectively
solves the problem in linear time. 

\subsection{The optimization problem\label{sub:opt}}
Given the set of gNBs $\Gc$, $N$, the number of supported beams at
each gNB, and  the set of zones $\Zc$, our aim is to jointly address
the two following questions while maximizing the overall achieved
network rate: i) what beam design should each gNB employ, i.e., how many beams, of what width, and direction; and  ii) which zones should be associated to which gNB and scheduled on which beam. 

To this end, we formulate an optimization problem that needs to be
solved periodically at every time step $k\in\Kc$. Given the solution,
this is provided to the set of gNBs, which update their beams design
accordingly. 
Since the problem formulation holds at every time step, to simplify
the notation, in the following we drop the time index $k$. 

Let us first define a set of beams $\Bc$ available at every gNB. 
Each beam $b\in\Bc$ is defined by a direction $\delta_b$ and
half-power beamwidth $\alpha_b$. 
If we consider a finite set of possible directions $D$  and a finite
set of possible beamwidths $A$, then the set of beams, $\Bc$, at the gNB  would contain $|D||A|$ potential beams.  

 Let $\pi(g,b)$ be the binary variable indicating
whether beam $b$ at gNB $g$ is employed or not, and let
$\gamma(g,b,z)$ be the binary variable indicating that zone $z$ is
associated with gNB $g$ on beam $b$.  
To assess whether a beam $b$ at gNB $g$ can cover a zone $z$, we
proceed as follows. 
First, for each zone $z$, we derive geometrically the LoS direction from a
gNB $g$, i.e., the angle of departure, denoted as
$\theta(g,z)$\footnote{All directions are defined in reference to a
  global coordinate system.}. 
We then  denote with $\Cc_z$ the   set of all $(g,b)$ tuples
covering zone $z$, i.e., all tuples for which $\pi(g,b)=1$ and which
fulfil the condition\footnote{Recall that a zone size is negligible
  with respect to any beam footprint.} $|\theta(g,z)-\delta_b|\leq\frac{\alpha_b}{2}$.

The optimization problem is then defined as follows:
\begin{equation}\label{eq:opt-obj}
\max_{\bm{\pi},\bm{\gamma}}\sum_{g\in\Gc}\sum_{b\in\Bc}\sum_{z\in\Zc} \pi(g,b)\gamma(g,b,z)R(g,b,z)
\end{equation}
where $ R(g,b,z)$ is the achieveable rate at zone $z$ from gNB $g$ on beam $b$, given the set of indicator variables $\bm{\pi}$ and $\bm{\gamma}$.

 The achieveable rate at zone $z$ is given by the following expression:
\begin{multline}\label{eq:rate}
R(g,b,z)\mathord{=}W\sum_{v\in\Vc}\log_2\left(1 +\frac{P(g,b)\left|\tilde{h}(g,b,v)\right|^2}{N_0+I_v}\right)
\end{multline}
where $W$ is the system bandwidth, while the second term within the
logarithmic function is the signal-to-interference-and-noise (SINR)
ratio. 
In the numerator, $\tilde{h}_c(g,b,v)$ represents the channel gain between gNB $g$ and
the vehicles $v$ in zone $z$ on beam $b$, while $P(g,b)$ is the
power allocated to beam $b$ by $g$. At the denominator,
$N_0$ represents the white noise power, while $I_v$ represents the
interference experienced by the vehicle in the zone from all other
active $(g',b')$ tuples in $\Cc_z$  to which $z$ is not associated
with.
  For any vehicle $v$ in zone $z$, $I_v$ can be expressed as:
\begin{equation}
I_v\mathord{=}\sum_{(g',b')\in \Cc_z}[1-\gamma(g',b',z)]\pi(g',b')P(g',b')|\tilde{h}(g',b',v)|^2\,.
\end{equation}
The channel gains, $\tilde{h}_c(g,b,v)$ which account for propagation losses and the beamforming gains, are derived according to \cite{zorzi-channel}.  

Next, we  define the constraints characterizing the system under
study. First, we limit the number of simultaneous beams that can be used by the gNB: 
\begin{equation}
\sum_{b\in\Bc}\pi(g,b)\leq N, \forall g\in \Gc \,.
\end{equation}
gNBs must also adhere to a power budget and ensure that power is not allocated to unused beams, $P^t(g)$, namely:
\begin{equation}
\sum_{b\in\Bc}P(g,b)\leq P^t(g), \forall g\in \Gc\,.
\end{equation}
\begin{equation}
P(g,b)\leq \pi(g,b) P^t(g), \forall g\in \Gc, b\in\Bc \,.
\end{equation}
Finally, we must  ensure that beams do not overlap with each other, i.e., for every two beams $b_i, b_j$ at $g$, for which $\pi(g,b)=1$, the following condition must hold:
\begin{equation}
|\delta_{b_i}-\delta_{b_j}|\geq\frac{\alpha_{b_i}+\alpha_{b_j}}{2},
\end{equation}

 We then impose constraints on the receiving end of the
 communication. 
First, we ensure that no zone $z$ is associated with a $(g,b)$ tuple that
cannot cover that zone:
\begin{equation}
\sum_{(g,b)\notin \Cc_z}\gamma(g,b,z)\leq 0, \forall z\in\Zc\,,
\end{equation}
and that no zone $z$ is scheduled on an inactive beam: 
\begin{equation}
\gamma(g,b,z)\leq\pi(g,b), \forall g\in\Gc,b\in\Bc, z\in\Zc\,.
\end{equation}

In addition, 
for CoMP-like communications, in which several gNBs coordinate to
transmit the same data to a certain zone, we impose that:
\begin{equation}
\sum_{g,b\in c_z}\gamma(g,b,z)\leq L, \forall z\in\Zc,
\end{equation}
where $L$ is the maximum number of gNBs that can partake in the
coordinated tranmission.   
Clearly, when no CoMP is enabled, $L=1$. 

The problem contanis nonlinear equations, e.g., \Eq{rate}, and contains
integer variables, namely,~$\pi$ and~$\gamma$; it therefore falls into
the category of nonlinear integer
programming~\cite{hemmecke2010nonlinear}. Such problems are more complex
than mixed-integer linear programming (MILP) problems, which are
themselves known to be NP-hard~\cite{boyd}. 
While there are algorithms that do solve MILP problems to the
optimum, 
solution strategies for non-linear integer problems only find
local optima in the general case.

\subsection{A graph-based approach\label{sub:graph}}

To overcome the complexity of the above problem, we proceed as
follows.
First,  we develop a graph-based
model of the system that captures all the essential aspects of the
mmwave vehicular networks. Second, we leverage  such a model to devise
an effective, 
linear-complexity  heuristic.

\begin{figure}
\centering
\includegraphics[width=.9\columnwidth]{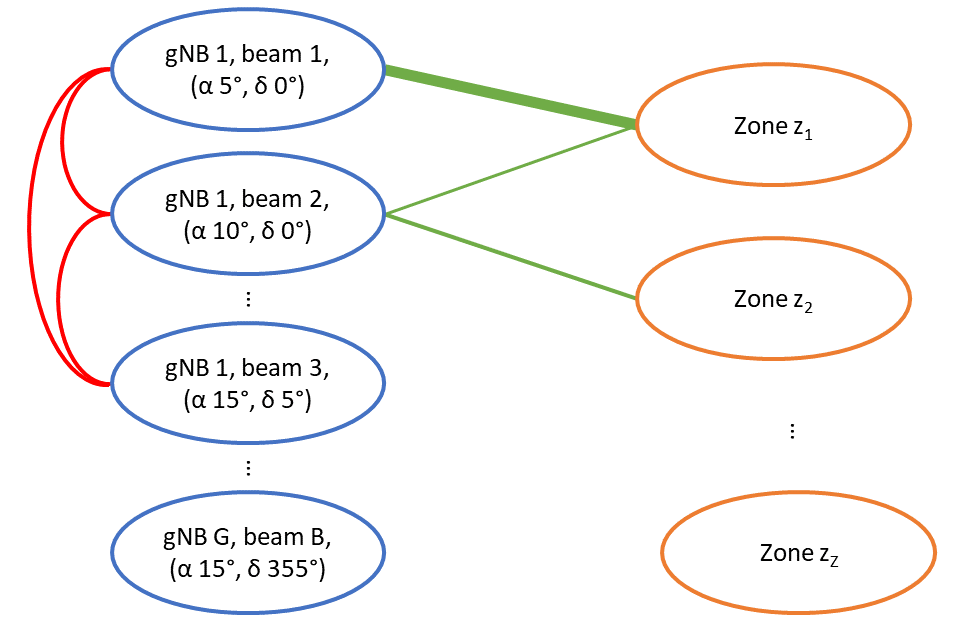}
\caption{\label{fig:bipartite}
    Example of bipartite graph. Left-hand, blue vertices represent beam steering decisions; right-hand, orange vertices represent zones. Green edges represent coverage opportunities, and their weights represent the corresponding rate. Red conflict edges join mutually-exclusive options.
} 
\end{figure}

As already mentioned, an effective beam design requires to
{\em{jointly}} set (i) 
 the number of beams of each gNB,
(ii) their width, and 
(iii) their direction,   
while respecting the system constraints.
Making these decisions sequentially, e.g., first deciding the width of beams and then their directions, is possible but likely to result in suboptimal solutions. On the other hand, making them jointly requires recognizing and accounting for the {\em conflicts} between decisions, i.e., the fact that several options are mutually exclusive and taking one renders the others invalid. 
To this end, we first cast the task of beam design and user association into a problem of {\em bipartite weighted matching with conflicts}. Specifically, we build a bipartite graph similar to the one in \Fig{bipartite}, where:
\begin{itemize}
    \item left-hand side vertices represent $(g,b)$ tuples;
    \item right-hand side vertices represent zones;
    \item edges between left- and right-hand side vertices represent
      the fact that a certain beam covers a certain zone, i.e., that
      $(g,b)\in \Cc_z$;
\item the edge weights correspond to the achievable rate $R(g,b,z)$ between left-hand vertex $(g,b)$ and right-hand vertex $z$;
    \item conflict edges, drawn between left-hand side vertices, denote combinations that are mutually exclusive.
\end{itemize}
With reference to the example in \Fig{bipartite}, we can observe that
conflict edges are drawn between vertices corresponding to the same
gNB, with incompatible beam choices, e.g., combinations of decisions
that would result in beams of the same gNB overlapping. To make the
matching problem tractable, the weights of the edges, i.e., the
achievable rates, are calculated by taking into account only the
noise-limited rate, which is a fair assumption considering that mmwave networks have in general very limited interference. The selection of an edge between the left-hand and
right-hand  vertices corresponds to setting both binary variables $\pi(g,b)=\gamma(g,b,z)=1$.

Although the problem of weighted bipartite matching {\em with
  conflicts} is NP-hard~\cite{chen2016conflict}, some heuristic
algorithms are available in the literature, which we can be used for
an efficient and effective solution of the beam design problem.
Specifically, we leverage the algorithm presented in~\cite[Sec.~4.3]{chen2016conflict}, which operates in linear time, at the cost of a linear competitive ratio.


%

\section{Numerical results}
\label{sec:results}
The performance of our approach is evaluated through numerical
simulations of an urban vehicular network constructed through the
publicly available traces for the city of Luxembourg as described in
Sec.~\ref{sec:system-trace}. 

We limit our scenario to a 4\,km$^2$ area of the city, covering most
of its centre, 
in which 51 gNBs are distributed as depcited in
Fig.~\ref{fig:scenario}. The system parameters are as follows. The
central frequency is set at $f_c=76$\,GHz \cite{comm-radar}, while the
bandwidth is set at $W=400$\,MHz as foreseen in  5G networks
\cite{zorzi-tutorial}. All gNBs are equipped with $32\times32$ UPA, with antenna elements spaced by $\lambda/2$, that
can transmit up to $N=4$ beams simultaneously. The transmit power is
limited to $P^t=30$\,dBm for any gNB. The vehicles are equipped with
$8\times8$ UPA, and can receive on a single beam at a time, which is
always directed towards the associated gNB. The composite effect of
channel and beamforming gains are modeled in accordance with
\cite{zorzi-channel}, as well as LoS and outage probability which are tailored to the Luxembourg
scenario. 
Furthermore, we consider three supported beamwidths $A=\{5^{\circ},
10^{\circ}, 15^{\circ}\}$, while beam directions can be any integer
number between $0^{\circ}$ and $359^{\circ}$. 

We compare the performance of the proposed approach against a
clustering-based technique, using the low-complexity but efficient
DBSCAN algorithm \cite{dbscan} to generate the number and directions
of the beams. This is used as a benchmark approach. It should be noted
that DBSCAN cannot determine the width of the beams,
therefore, we try all possible values and take the one resulting in the best performance.
Both algorithms are executed periodically
every 1~s, and the total simulation duration is 20~s. The underlying
allocation of resources is performed using the Proportional-Fair
scheduling algorithm. In the following plots, CAWBM denotes the
proposed conflict-aware weighted bi-partite matching approach.

\begin{figure}
\centering
\includegraphics[width=0.4\textwidth]{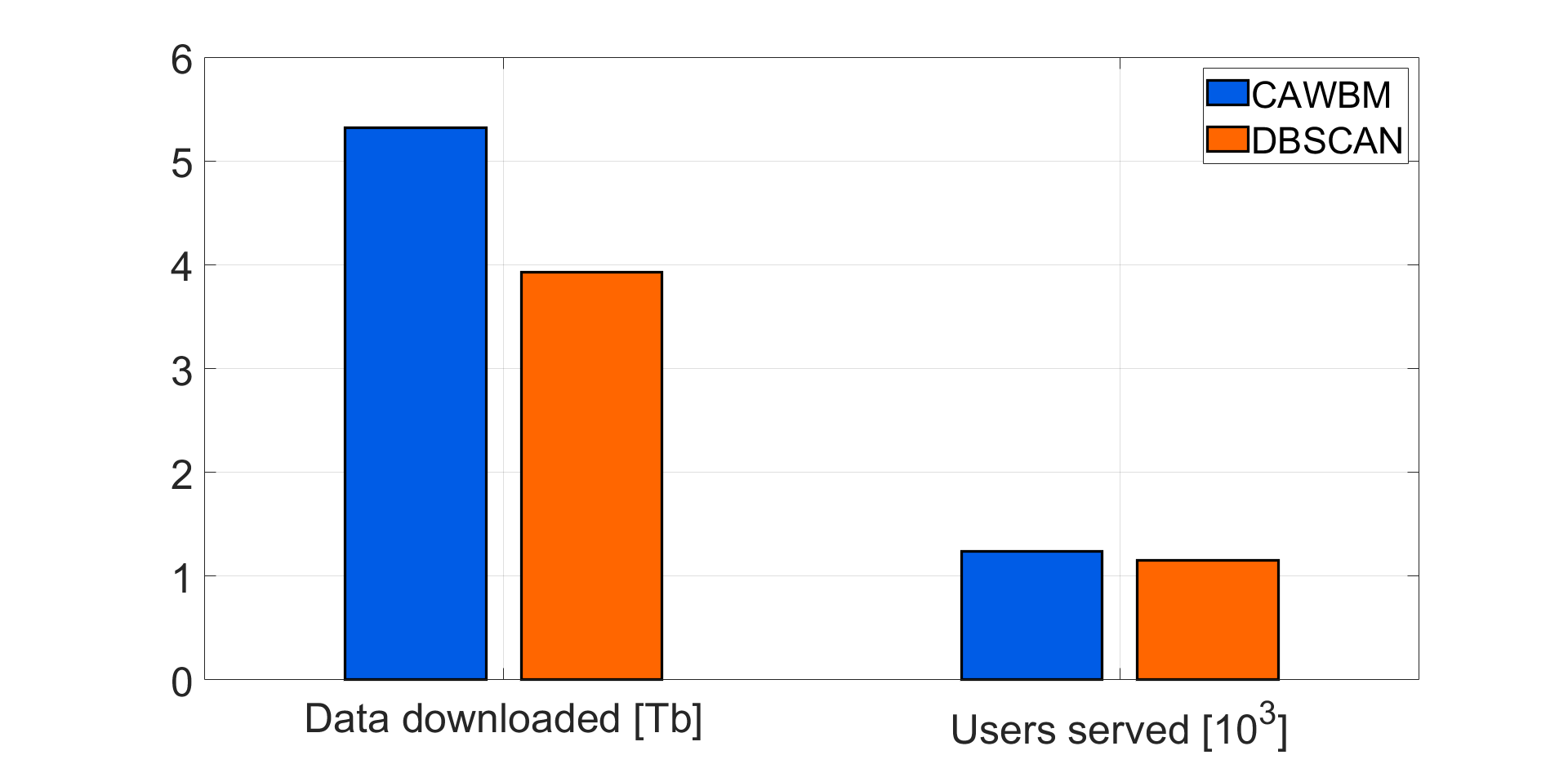}
\caption{\label{fig:totaldata} Total amount of data downloaded [Tb] and number of users served [$10^3$]. }
\end{figure}
We first look at the perfomance of both approaches in absolute terms, by evaluating the total number of users served over the 20-s period and the amount of data downloaded, as shown in Fig.~\Fig{totaldata}. CAWBM can serve around 35\%, roughly 1.4 Tb, more data than DBSCAN, and around 7\% more users. In particular, CAWBM serves 92\% of all vehicles in the network, while DBSCAN serves around 85\% of them.

\begin{figure}
\centering
\includegraphics[width=.23\textwidth]{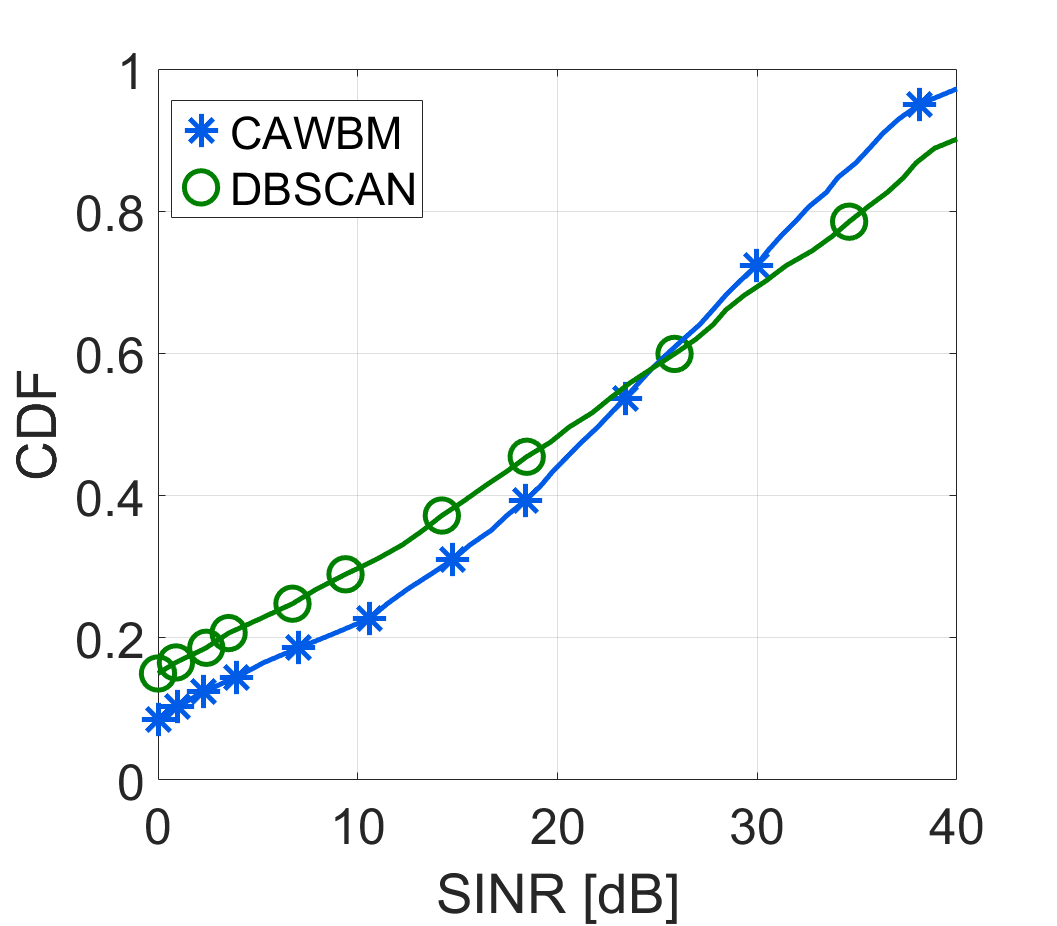}
\includegraphics[width=.23\textwidth]{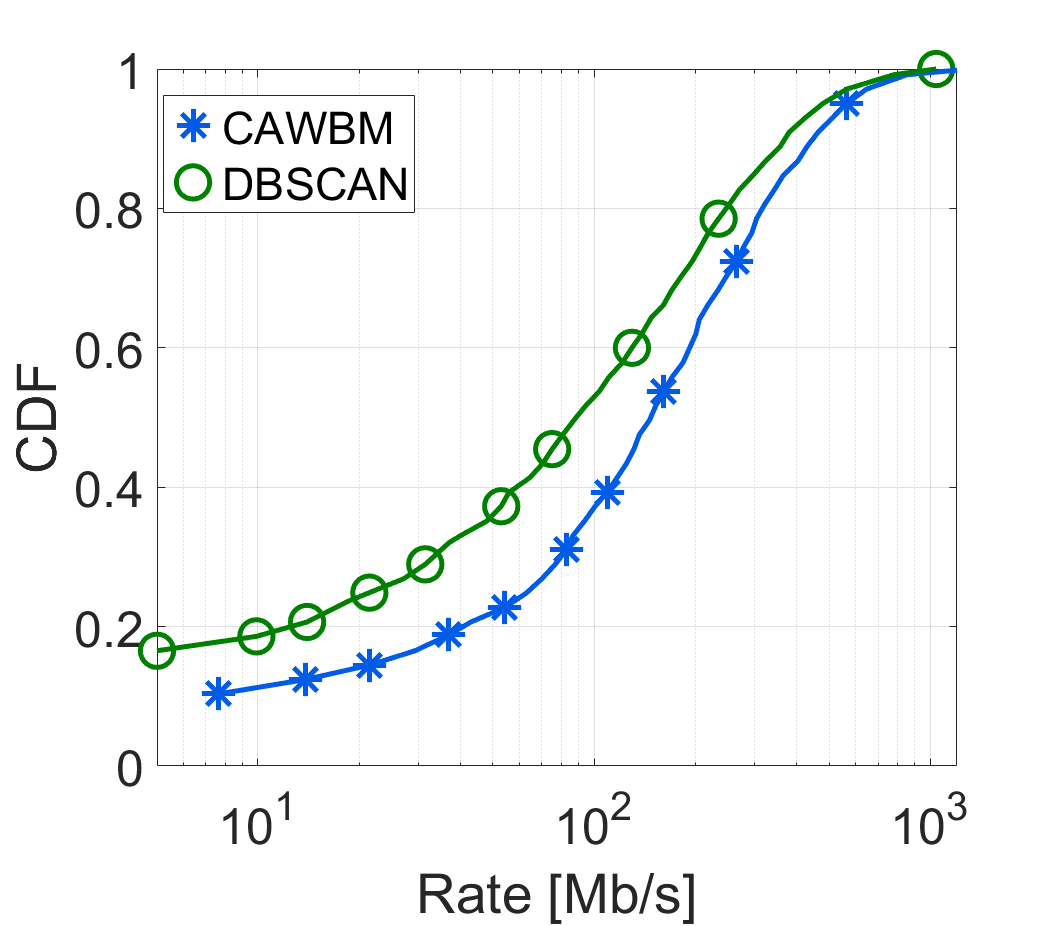}
\caption{CDF of average SINR experienced by vehicles (left) and effective rate experienced by served vehicles (right).
    \label{fig:rate}
} 
\end{figure}

Next, we look at the cummulative distribution function (CDF) of the
experienced average SINR and  data rates achieved by the
vehicles. During the simulations, the SINR was calculated taking into
account both the noise and interference, and then mapped into the data
rate by using the 4-bit channel quality indicator (CQI) table in
\cite{3gpp-tech}. It is interesting to note that while DBSCAN can
ensure slightly better SINR values for the top 40\% of the users, as
seen in \Fig{rate}(left), this behavior is not reflected in the achieved rates by the vehicles, shown in \Fig{rate}(right). 

\begin{figure}
\centering
\includegraphics[width=.23\textwidth]{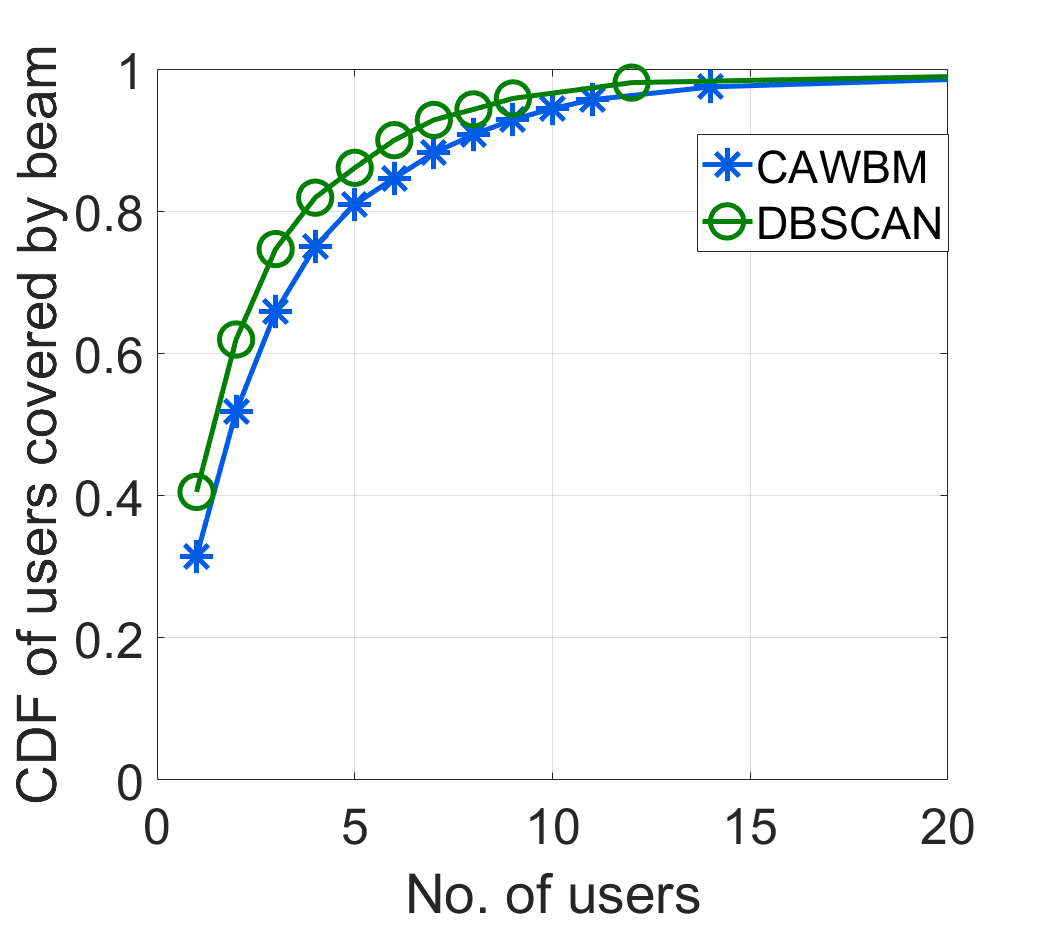}
\includegraphics[width=.23\textwidth]{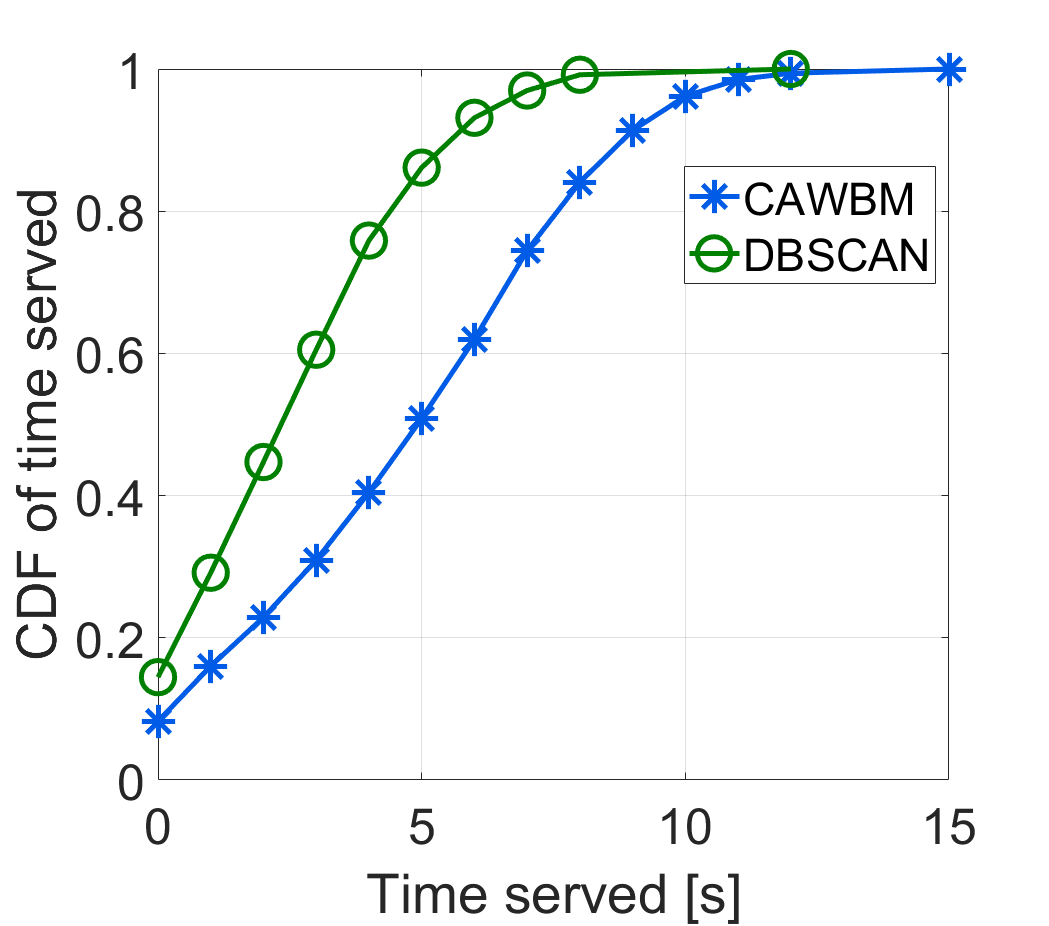}
\caption{CDF of number of users served by each beam (left) and CDF of amount of time each vehicle is served (right).
    \label{fig:beamtime}
} 
\end{figure}

This can be explained by the fact that the CAWBM approach serves both more vehicles per beam (as shown in \Fig{beamtime}(left)) and for a longer period of time (\Fig{beamtime}(right)).  
The improvement in performance brought by CAWBM when compared to
DBSCAN can be attributed to the fact that the former takes into
account the rates of all potential vehicles within the small area of
the zone. DBSCAN, on the other hand, acts only on the information
regarding the vehicle relative LoS direction towards the gNBs. In
addition, CAWBM is likely to favor beams that cover zones that are
both more frequented and well positioned to experience higher levels
of SINR. This is the reason why over 50\% of the vehicles are served
more than 5~s with CAWBM, while, under DBSCAN, the corresponding
percentage of vehicles is  just 15\%. 

\section{Conclusions and future work}
\label{sec:conclusion}
While mmwave communications have emerged as a promising candidate technology
for future vehicular networks, the performance of mmwave networks heavily depends upon beam management aspects.  
The need for adequate
alignment of beams between gNBs and vehicles is critical, and, as such,
 efficient beam design becomes paramount. We adressed both beam design
 aspects and user association though a graph-based approach. Once we
 modeled our system as a weighted bipartite graph, we were able to
 cast the problem at hand as a conflict-aware matching problem, which
 can be efficiently solved in linear time, through heuristic algorithms.   Our performance evaluation, based on real-world
topology and mobility information, has provided relevant
insights. Thanks to the conflict-aware approach, the solution we
proposed significantly outperforms our benchmark scheme  leveraging a clustering algorithm.

Future work will focus on further improving the mmWave graph
model, and further investigating the interaction between gNBs during
the beam design phase.

\section*{Acknowledgement}
This work has been performed in the framework of the European Union’s
Horizon 2020 project 5G-CARMEN co-funded by the EU under grant
agreement No.\, 825012, and has also been partially supported by the Academy of Arts and
Sciences of Kosovo.

\bibliography{mmwave}
\bibliographystyle{IEEEtran}
\end{document}